\documentstyle[12pt,epsf]{article}

\newcommand{\lab}[1]{\label{#1}}

\newcommand{\rf}[1]{\ref{#1}}

\newcommand{\wt}{\widetilde}

\newcommand{\spc}{{\hspace{1em}}}

\newcommand{\filename}{\underline{\Large Filename : \tt\jobname.tex}}

\def\Tr{\mathop{\rm Tr}\nolimits}

\newcommand{\rap}[2]
{\setbox1=\hbox{#1}%
\setbox2=\hbox to\wd1{\hss #2\hss}%
\mbox{\rlap{\box1}\box2}}

\newcommand{\J}[4]{{\sl #1} {\bf #2} (#3) #4}
\newcommand{\NP}{Nucl.\ Phys.}
\newcommand{\PL}{Phys.\ Lett.}

\begin{document}

\baselineskip=18pt plus 0.2pt minus 0.1pt

\begin{titlepage}
\title{
\hfill\parbox{4cm}
{\normalsize KUNS-1436\\HE(TH)~97/03\\{\tt hep-th/9703077}}\\
\vspace{1cm}
A Comment on Fundamental Strings\\ in M(atrix) Theory
}
\author{
Yosuke Imamura\thanks{{\tt imamura@gauge.scphys.kyoto-u.ac.jp}}
{}\thanks{
Supported in part by Grant-in-Aid for Scientific
Research from Ministry of Education, Science and Culture
(\#5416).}
\\[7pt]
{\it Department of Physics, Kyoto University, Kyoto 606-01, Japan}
}
\date{}

\maketitle
\thispagestyle{empty}

\begin{abstract}
\normalsize
We present a solution of M(atrix) theory describing type IIA
fundamental string.
Our construction is based on the central charge of the longitudinal
membrane (= fundamental string), the BPS saturation condition
and the relation between M(atrix) theory and supersymmetric Yang-Mills
theory. The fundamental string corresponds to a photon in
supersymmetric Yang-Mills theory.
\end{abstract}

\end{titlepage}

\section{Introduction}
Recently Banks, Fischler, Shenker and Susskind \cite{BFSS} proposed
 M(atrix) model as a nonperturbative unified
description of M-theory \cite{WittenM} and all the string theories.
They claimed that the large $N$ limit of
a supersymmetric matrix quantum mechanics describing $N$ D-particles
is equivalent to M-theory.
To confirm this conjecture,
many nontrivial consistency checks were carried out to
attain desirable results.
For example,
it was shown \cite{BFSS} that
 the D2-brane tension and
 the long range force between moving D-particles
 coincide with the values expected from
supergravity arguments.
Similar checks were also done in \cite{LifschytzMathur,AharonyBerkooz}.
Other even dimensional objects, such as D4 and D6-brane,
were also constructed, and the potentials between them were calculated
to agree with string theory expectations\cite{Lifschytz}.

It is believed that in eleven dimensional supergravity
there are two dimensional object (M2-brane)
and five dimensional one (M5-brane),
which are coupled to the three-form field $A_{\alpha\beta\gamma}$
and its dual field $\wt A_{\mu_1\ldots\mu_6}$, respectively.
These objects correspond to D-objects, fundamental string and NS5-brane
in type IIA string theory.
The consistency of the correspondence was analyzed in detail in \cite{Alwis}.
Among these objects, D0, 2, 4-branes have been constructed in M(atrix) theory
\cite{BFSS,BFM}.
If M(atrix) theory describes all the states of M-theory as expected,
fundamental string (longitudinal membrane) and NS5-brane (transverse M5-brane)
must also be constructed in M(atrix) formulation.

As for transverse M5-brane,
Banks et al.\ \cite{BFM} analyzed the supersymmetry algebra
in M(atrix) theory and found that the central charge
 of transverse M5-brane is missing from the algebra.
This might imply the impossibility of constructing transverse M5-brane
in M(atrix) theory.
On the other hand, longitudinal M2-brane charge exists in the M(atrix)
theory supersymmetry algebra \cite{BFM} and it is constructed
from variables $X_a$ and $\theta$.
This fact strongly suggests the existence of longitudinal M2-brane
(fundamental string) solution of M(atrix) theory.

The purpose of this paper is to construct the solution describing the
fundamental string in M(atrix) theory.
To get the solution, we study the expression of the string central
charge in M(atrix) theory supersymmetry algebra\cite{BFM} and use the
BPS saturation condition.
It is found that the fundamental string
corresponds to the ``photon'' in supersymmetric Yang-Mills theory (SYM).

We shall comment on earlier works \cite{BFSS,Motl,SFM,Verlinde^2}
constructing fundamental string solution in M(atrix) theory.
In \cite{BFSS}, it was claimed that if we compactify one of the transverse
directions and T-dualize the theory along that direction,
a string world sheet action is obtained, which they identified
as that of type IIA fundamental string.
Because the one dimensional object
which is obtained by T-dualization of M(atrix) theory
is not a fundamental string but a D-string,
we need further S-dualization to identify it with the fundamental
string%
\footnote{In the theory thus obtained by T- plus S-dualizations of
        M(atrix) theory, all kinds of extended objects are
        described in terms of two dimensional $S_N$ orbifold sigma-model
        \cite{Motl,SFM,Verlinde^2}.
        This S-dualization prescription is equivalent
        to the exchange of two directions $x^9$ and $x^{11}$
        as is explained in \cite{Verlinde^2}.}.
%
%Therefore, this is different from the string studied in this paper.
The fundamental string solution we present in this paper is the one in
non-S-dualized description, and therefore is different from the
fundamental string discussed in \cite{BFSS,Motl,SFM,Verlinde^2}.
%In this paper we shall present the fundamental string solution in
%non-S-dualized description,
%
Indeed, the same objects are described
in quite different ways in each formulation.
For example, in the non-S-dualized description,
D-particles are represented by diagonal elements of matrices,
while from the viewpoint of S-dualized description
it is represented as an electric flux on the world sheet.

We shall add a comment on the longitudinal radius $R$.
If we want to reproduce (uncompactified) eleven dimensional theory,
we must take the limit $R\rightarrow\infty$.
However, because our purpose is to study fundamental string,
we should keep the string coupling constant weak, implying small $R$.
Therefore in this paper we use the words ``M-theory''
and ``M(atrix) theory'' as eleven dimensional theory
compactified on a small $S^1$, although they usually imply
a decompactified theory.

Finally, we present a few formulas used in later
sections.
In the M-theory context, ten dimensional type IIA theory with coupling
$g_{\rm str}$ is equivalent to eleven dimensional supergravity
compactified on $S^1(R)$ with radius $R$ related to the string coupling
by
\begin{equation}
R=g_{\rm str}^{\frac{2}{3}}l_s \ ,
\end{equation}
where $l_s$ is the string length scale
(we shall adopt the convention  $l_s=1$ hereafter).
Another important relation is the rescaling formula of metrics,
\begin{equation}
G^{(11)}_{\mu\nu}=g_{\rm str}^{-\frac{2}{3}}G^{(10)}_{\mu\nu}. \lab{metric}
\end{equation}
This equation is usually expressed as
 $l_{\rm planck}=g_{\rm str}^{-\frac{1}{3}}l_s$.

%%%%%%%%%%%%%%%%%%%%%%%%%%%%%%%%%%%%%%%%%%%%%%%%%%%%%%%%%%%%%%%%%%%%%%%%
\section{Central charges of branes}
Ten dimensional SYM theory
has 16 supersymmetry transformation parameters $\epsilon$.
In addition to them, 16 trivial fermionic symmetries
$\delta\theta=\wt\epsilon$ exist.
The idea of M(atrix) theory is to regard these 32 fermionic symmetries
\begin{equation}
\delta X_a=\frac{i}{2}\epsilon\gamma_a\theta,\spc
\delta\theta=-\frac{1}{2}D_tX_a\gamma^a\epsilon
             -\frac{1}{8\pi}[X_a,X_b]\gamma^{ab}\epsilon
             +\wt\epsilon,
\end{equation}
as a superspace representation
of the eleven dimensional supersymmetry transformations \cite{BFSS}.
We can construct supercharges for these transformations as follows \cite{BFM}.
\begin{eqnarray}
\epsilon Q&=&\frac{\sqrt{R}}{2}
            \Tr\Bigl(P_a\left(\epsilon\gamma^a\theta\right)\Bigr)
           +\frac{\sqrt{R}}{8\pi}\Tr\left([X_a,X_b]
           (\epsilon\gamma^{ab}\theta)\right), \lab{dynamicalQ}\\
\wt\epsilon\wt Q&=&\frac{1}{\sqrt{R}}\Tr\left(\wt\epsilon\theta\right).
\end{eqnarray}
If we omit the second term on the RHS of eq.\ (\rf{dynamicalQ}),
these supercharges reproduce
eleven dimensional ordinary (= no central charge) supersymmetry
algebra.
The omitted term can be regarded as contributions
 from objects carrying the central charges.

Banks et al.\cite{BFM} obtained the expressions of central charges
of extended objects except the one for transverse N5-brane
by calculating the anti-commutation relations of supercharges.
Their result is
\begin{eqnarray}
Z_b&=&-i\frac{R}{2\pi}\Tr\Bigl(P_a[X_a,X_b]\Bigr)
      -i\frac{R}{4\pi}\Tr\Bigl(\theta^\alpha[\theta^\alpha,X_b]\Bigr),
\lab{Za}\\
Z_{ab}&=&\frac{i}{4\pi}\Tr[X_a,X_b],\lab{Zab}\\
Z_{abcd}&=&\frac{R}{8\pi^2}\Tr X_{[a}X_bX_cX_{d]}.\lab{Zabcd}
\end{eqnarray}
These are the central charges of
longitudinal membrane, transverse membrane and longitudinal 5-brane,
respectively.
Their $R$-dependence is consistent with
 the fact that the corresponding brane is longitudinal or transverse.
To be precise, we have to compactify the transverse directions
in order for these central charges to be well-defined.

Let us pay attention to the one brane charge $Z_a$
related to longitudinal M2-brane.
In the type IIA picture, it is the wrapping number of fundamental string.
If a BPS saturated solution with nonzero $Z_a$ exists,
its energy is equal to $|Z_a|$.
Thus the following equation holds
(neglecting the fermionic part):
\begin{equation}
H=\frac{R}{2\pi}\Bigl|\Tr\left(P_a[X_a,X_b]\right)\Bigr|
 =\frac{1}{2\pi}\left|\Tr\left(\dot X_a[X_a,X_b]\right)\right|,\lab{H11}
\end{equation}
where we have adopted the eleven dimensional supergravity metric.
Translating eq.\ (\rf{H11}) to the string metric by using eq.(\rf{metric})
and rescaling the variables as
 $X_a\rightarrow g_{\rm str}^{-1/3}X_a$, $t\rightarrow g_{\rm str}^{-1/3}t$,
$H\rightarrow g_{\rm str}^{1/3}H$, we obtain
\begin{equation}
H=\frac{1}{2\pi g_{\rm str}}\left|\Tr\left(\dot X_a[X_a,X_b]\right)\right|.
\lab{HinSTR}
\end{equation}
%%%%%%%%%%%%%%%%%%%%%%%%%%%%%%%%%%%%%%%%%%%%%%%%%%%%%%%%%%%%%%%%%%
\section{Relation to Supersymmetric Yang-Mills theory}
In \cite{WT4a,WT4b} it was shown that M(atrix) theory compactified
 on a torus $T^p$ with radius $R_a$ $(1\leq a \leq p)$
is equivalent
to $p+1$ dimensional SYM theory
on the dual torus $\wt T^p$ with radius $\wt R_a=1/R_a$
 via T-duality transformation.
A D-particle in IIA theory is translated to background $p$-brane
wrapped around $\wt T^p$,
and the variable $X_a$ to $U(N)$ gauge field $A_a$.

Compactification means that $X_a$ and $X_a+2\pi R_a$ are the same point,
and this fact is translated to that $A_a$ and $A_a+1/\wt R_a$ are
gauge equivalent by a large gauge transformation $\exp(ix_a/\wt R_a)$.
Thus a relation between $X_a$ and $A_a$ is
\begin{equation}
X_a=2\pi\nabla_a\equiv2\pi\left(\partial_a+A_a\right).\lab{X=A}
\end{equation}
Therefore, the membrane wrapping number $[X_a,X_b]$
and the velocity $\dot X_a$ in M(atrix) theory
correspond to the magnetic flux
$(2\pi)^2B_{ab}=[2\pi\nabla_a,2\pi\nabla_b]$
and the electric flux $2\pi E_a=2\pi\dot A_a$ in SYM, respectively.
The relation between the coupling constants
 $g_{\rm str}$ and $g_{\rm YM}$ can be derived
by comparing the two actions in the $10-p$ uncompactified dimensions:
\begin{equation}
\frac{(2\pi)^2}{g_{\rm str}}=\frac{V}{g_{\rm YM}^2},
\end{equation}
where $V$ is the volume of the dual torus $\wt T^p$.
%%%%%%%%%%%%%%%%%%%%%%%%%%%%%%%%%%%%%%%%%%%%%%%%%%%%%%%%%%%%%%%%%%
\section{Fundamental string solution}
Since the following argument does not depend essentially on $p$,
we shall consider the most familiar case $p=3$.
In this case, the magnetic flux is represented by a 3-vector
 $B_a=\frac{1}{2}\epsilon_{abc}B_{bc}$.
Rewriting eq.\ (\rf{HinSTR}) by using the correspondence (\rf{X=A}),
we have
\begin{equation}
H=\frac{1}{g_{\rm YM}^2}\left|\int dV\Tr\left(E\times
B\right)\right|.\lab{H=pv}
\end{equation}
In deriving this equation, we have used the expression of the central charge
(\rf{Za}) and the BPS saturation condition $H=|Z_a|$.

On the other hand, if we regard a configuration carrying nonzero
 $Z_a$ as a string, its energy is given as a product of
string tension $1/2\pi$ and the circumference $2\pi R_3$,
\begin{equation}
H=R_3=\frac{1}{\wt R_3}.\lab{H=1/R}
\end{equation}
(We assume that the winding is along the 3rd direction, which implies that
$Z_3\ne 0$. In the M-theory view point, we are considering the
M2-brane extended in the longitudinal and $3$rd directions.)
Combining eqs.\ (\rf{H=pv}) and (\rf{H=1/R}), we obtain
the key equation which the fundamental string configuration should
satisfy:
\begin{equation}
\frac{1}{\wt R_3}=\frac{1}{g_{\rm YM}^2}
      \left|\int dV\Tr\left(E\times B\right)\right|.
\lab{kk=pv}
\end{equation}
This equation is interpreted as follows.
The LHS is one unit of Kaluza-Klein momentum
and the RHS is the Poynting vector of Yang-Mills field.
Therefore, eq.\ (\rf{kk=pv}) tells us that the configuration
corresponding to fundamental string wrapped around the $3$rd direction
is a plane wave of one photon moving along the $3$rd direction.
To be precise,
unless the number of photons is sufficiently large,
it is impossible to determine
the field strength ($E$ and $B$) and the photon momentum
simultaneously owing to the uncertainty principle.
Here we shall ignore this problem and
consider a classical plane wave with Poynting vector $1/\wt R_3$.

First let us consider the plane wave taking values
in the $U(1)$ part of $U(N)$.
The field strength of such a plane wave is
\begin{equation}
E_1=B_2=\frac{g}{2\pi^{3/2}\sqrt{\wt R_1\wt R_2}\wt R_3}
                   \sin\frac{x_3-t}{\wt R_3}H_i
                   ,\spc \mbox{others}=0,
\end{equation}
where $(H_i)_{mn}=\delta_{im}\delta_{mn}$ $(i=1,\ldots,N)$ are the
generators of the $U(1)$ part.
The corresponding vector potential is given by
\begin{equation}
A_1=c\cos\frac{x_3-t}{\wt R_3}H_i,\spc
A_2=0,\spc
A_3=\sum_k\frac{a_k}{\wt R_3}H_k.\spc
c=\frac{g}{2\pi^{3/2}\sqrt{\wt R_1\wt R_2}},\spc
\lab{Ais}
\end{equation}
where $a_k$ are Wilson lines along the 3rd direction.
For simplicity, we have assumed that $A_1$ and $A_2$ have no
static parts, implying that all the D-particles sit on the $x^3$
axis.
This result looks natural
if we note that type IIA string theory
and SYM theory
are T-dual with one another,
and under this duality wrapped string in type IIA theory is transformed to
the momentum mode of the open string, i.e., photon on the D-brane.

Since we have found the vector potential $A_a$
 representing the fundamental string,
our next task is to obtain the matrix representation $X_a$
of the covariant derivative (\rf{X=A}).
Taking the basis
$\phi_m(n_1,n_2,n_3)\equiv
\exp\left(i\sum\limits_{i=0}^3n_ix_i/\wt R_i\right)e_m$,
where $e_m=(0,\ldots,1,\ldots,0)^T$ is a unit vector whose $m$th component is
$1$,
we obtain
\begin{eqnarray}
&&\nabla_1\vec\phi\left(n_1,n_2,n_3\right)
=\frac{n_1}{\wt R_1}\vec\phi\left(n_1,n_2,n_3\right)\nonumber\\
&&\hspace{9em}
   +\frac{c}{2}e^{-i\frac{t}{\wt R_3}}
               H_i\vec\phi\left(n_1,n_2,n_3+1\right)
   +\frac{c}{2}e^{+i\frac{t}{\wt R_3}}
               H_i\vec\phi\left(n_1,n_2,n_3-1\right) ,
           \nonumber\\
&&\nabla_2\vec\phi\left(n_1,n_2,n_3\right)
=\frac{n_2}{\wt R_2}\vec\phi\left(n_1,n_2,n_3\right) ,\nonumber\\
&&\nabla_3\vec\phi\left(n_1,n_2,n_3\right)
=\frac{1}{\wt
R_3}\left(n_3+\sum_ka_kH_k\right)\vec\phi\left(n_1,n_2,n_3\right).
\end{eqnarray}
The first terms on the RHS of these equations
 exist even in the case where photon is absent.
They correspond to background D3-brane (or D-particle before T-dualization).
Therefore the components which represent the fundamental string
(photon) are
\begin{equation}
(X_1)_{(n_1,n_2,n_3),(n_1,n_2,n_3\pm1)}
=\pi cH_i\exp\left(\mp i\frac{t}{\wt R_3}\right).
\lab{solution}
\end{equation}
The oscillation in time is due to the energy of the photon.

The correspondence between fluctuations of non-diagonal part of $X_a$
and an open string stretched between two D-particles is
a fundamental concept in M(atrix) theory.
So far this non-diagonal part has been treated
only as a virtual string which contributes to a potential
between D-objects.
We emphasize that our solution (\rf{solution}) represents
not the virtual string but the static real BPS string.

Eq.\ (\rf{solution})
represents a wrapped string whose both the ends are on
the same D-particle $i$ (A in Fig.\ \rf{3string}).
In this case, the photon is neutral for all the $U(1)$s,
which are the remnants of $U(N)$ gauge symmetry
(we are considering the generic case where gauge symmetry is broken to
$U(1)^N$).
In addition to these $N$ neutral photons,
$N^2-N$ charged particles, which correspond to the non-diagonal part
of $U(N)$,
also exist.
We call them ``W-boson''.
Every W-boson is charged with respect to two of the $N$ $U(1)$s.
These W-bosons represent a string whose ends are on different
D-particles corresponding to the two $U(1)$s (B in Fig.\ \rf{3string}).
However this configuration is
unstable and cannot be an exact solution of M(atrix) theory,
since the string shrinks due to its tension and the D-particles cannot
stand still.
To make it stable, we must add another string to pull the D-particles
to the opposite direction (C in Fig.\ \rf{3string}).
In SYM theory, this corresponds to the fact that a
single W-boson cannot exist in a compact space ($T^3$ in this case)
because flux has nowhere to go,
while two W-bosons whose charges are opposite each other can exist
(these W-bosons correspond to two strings of C in Fig.\ \rf{3string}).

\begin{figure}[hbt]
\epsfxsize=8cm
\begin{center}
\leavevmode
\epsfbox{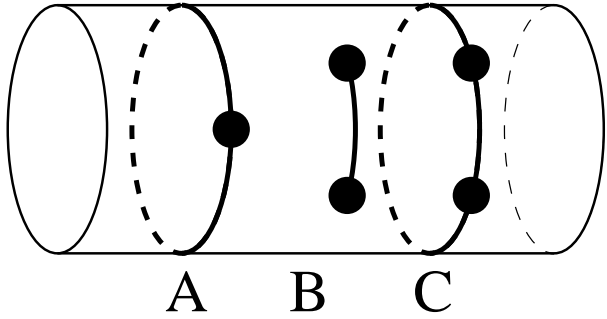}
%\\[2ex]{\LARGE Figure \rf{3string}}
\caption{Wrapped string corresponding to a neutral photon (A),
 unstable configuration corresponding to a single charged W-boson (B)
 and stable configuration corresponding to two W-bosons
 whose charges are canceled out (C).
 The blobs represent D-particles.}
\lab{3string}
\end{center}
\end{figure}

In general, there exist stable states with arbitrary number of W-bosons.
For such states
we can prove the equivalence between ``the condition of balance of the
tension'' and ``the charge vanishing condition'' as follows.
If the tensions of the strings attached to each D-particle are balanced,
the energy of the system is stationary
against small shifts of the D-particles.
In SYM theory, this corresponds to the fact that
small changes of the Wilson lines
keep the energy unchanged.
The energy of this system is the sum of the energies $E_i$ of the
W-bosons (index $I$ labels the W-bosons).
They are given by
\begin{equation}
E_I=4\pi^2 R_3\Bigl|\sum_kq_{Ik}a_k+n_I\Bigr|,
\lab{E_i}
\end{equation}
where $q_{Ik}$ and $n_I$ are the charges of W-bosons
and the integral numbers labeling the Kaluza-Klein modes,
respectively, and index $k$ specifies one of the $N$ $U(1)$s.
The signature of the quantity inside the absolute value on the RHS of
(\rf{E_i}) corresponds to the winding directions of the string.
Because the BPS saturation condition demands all these direction to be the
same,
we can ignore the absolute value symbol.
Hence, the change of the total energy $E_{\rm total}=\sum_I E_I$
caused by the shift of the Wilson lines is
\begin{equation}
\frac{\partial E_{\rm total}}{\partial a_k}=4\pi^2R_3\sum_Iq_{Ik}.
\end{equation}
Therefore, the stationarity of the total energy
means that the sum of the charges of W-bosons vanishes.

Taking these points into account,
we can easily construct exact solutions of the equation of motion
$\ddot X_a+[X_b,[X_b,X_a]]/(2\pi)^2=0$
as a sum of the oscillation modes representing each W-boson:
\begin{eqnarray}
&&X_1=X^{(0)}_1+\sum_{I=1}^MX^{(I)}_1,\spc\nonumber\\
&&(X_2)_{(n_1,n_2,n_3),(n_1,n_2,n_3)}=2\pi R_2n_2,\spc\nonumber\\
&&(X_3)_{(n_1,n_2,n_3),(n_1,n_2,n_3)}=2\pi R_3
\left(n_3+\sum_ka_kH_k\right),
\lab{General}
\end{eqnarray}
where
\begin{eqnarray}
&&(X^{(0)}_1)_{(n_1,n_2,n_3),(n_1,n_2,n_3)}=2\pi R_1n_1,\nonumber\\
&&(X^{(I)}_1)_{(n_1,n_2,n_3),(n_1,n_2,n_3)}
=\pi c\left(
   \exp(+i(a_{i_{I+1}}-a_{i_I})t)E_{i_Ii_{I+1}}+\rm{h.c.}
  \right) \spc\mbox{for $1\leq I\leq M-1$},\nonumber\\
&&(X^{(M)}_1)_{(n_1,n_2,n_3),(n_1+1,n_2,n_3)}
=\pi c\exp(+i(a_{i_1}+1-a_{i_M})t)E_{i_Mi_1},\nonumber\\
&&(X^{(M)}_1)_{(n_1+1,n_2,n_3),(n_1,n_2,n_3)}
=\pi c\exp(-i(a_{i_1}+1-a_{i_M})t)E_{i_1i_M}.
\lab{where}
\end{eqnarray}
In eqs.\ (\rf{General}) and (\rf{where}),
$i_I$ and $i_{I+1}$ are the numbers which satisfy
$q_{Ii_I}=+1$ and $q_{Ii_{I+1}}=-1$,
$M$ is the total number of the W-bosons,
and we have assumed that $0\leq a_{i_1}<a_{i_2}<\cdots<a_{i_M}<1$.

%%%%%%%%%%%%%%%%%%%%%%%%%%%%%%%%%%%%%%%%%%%%%%%%%%%%%%%
\section{Conclusion}
In this paper we have obtained a solution of M(atrix) theory
corresponding to fundamental string.
It is given as an oscillation mode of non-diagonal part of matrices
$X_a$, eq.\ (\rf{solution}), and represents a static BPS string.

There are a number of questions to be clarified.
For example, we do not know how to reproduce the following three
in M(atrix) theory:
\begin{itemize}
\item
Infinite tower of massive modes of string.
\item
Non-wrapping closed string, in particular, graviton.
\item
Winding BPS states of closed string.
\end{itemize}
We shall comment on the first and the last objects.

Concerning the infinite tower of massive modes of string, we should
comment on \cite{Motl,SFM,Verlinde^2}.
Although they reproduce the massive modes of string, the massive modes are the
ones of the string in the S-dualized description (recall the
explanation in Sec.\ 1).
What we would like to reproduce here is the massive modes
in non-S-dualized description, which correspond to the excitation
modes of D-objects in the S-dualized description
(however, they are not discussed in \cite{Motl,SFM,Verlinde^2}).

Next we shall comment on the winding closed string.
If a solution representing
 the winding closed string (B in Fig.\ \rf{oandc}) exists,
 it must satisfy eq.\ (\rf{kk=pv}) like the open string states we have
discussed (A in Fig.\ \rf{oandc}).
However, every element $X_{mn}$ corresponds to the string stretched between
D-particles.
So within our framework there is no way to describe winding closed strings
 which are not connected to D-particles.
Two types of states (A and B in Fig.\ \rf{oandc}) have the same energy,
and transition between them must exist to obey unitarity,
so that the winding closed string state should also exist in the
theory.

\begin{figure}[hbt]
\epsfxsize=8cm
\begin{center}
\leavevmode
\epsfbox{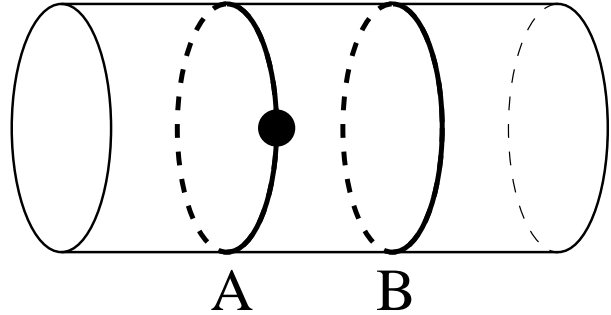}
%\\[2ex]{\LARGE Figure \rf{oandc}}
\caption{An open string (A) pinned to a D-particle (blob) and a
 winding closed string (B).}
\lab{oandc}
\end{center}
\end{figure}

\vspace{5ex}
I would like to thank H.\ Hata and T.\ Kugo
for valuable discussions and careful reading of the manuscript.

\vspace{2ex}
\noindent
{\bf Note added:} After completing this paper, I became aware of
ref.\,\cite{IIBM}, which gives a similar argument as mine.
I would like to thank Pei-Ming Ho for informing me of their work.

%%%%%%%%%%%%%%%%%%%%%%%%%%%%%%%%%%%%%%%%%%%%%%%%%%%%%%%
%%%%%%%%%%%%%%%%%%%%%%%%%%%%%%%%%%%%%%%%%%%%%%%%%%%%%%%

%%%%%%%%%%%%%%%%%%%%%%%%%%%%%%%%%%%%%%%%%%%%%%%%%%%%%%%%%%%%%%%%%%%%%%%
\end{document}